\documentclass[aps,showpacs,twocolumn]{revtex4}
\usepackage{epsfig}
\usepackage{color}
\usepackage{amsmath}
\usepackage{amssymb}
\usepackage{bbm}

\newcommand{\be}{\begin{equation}}
\newcommand{\ee}{\end{equation}}
\newcommand{\bea}{\begin{eqnarray}}
\newcommand{\beaa}{\begin{eqnarray*}}
\newcommand{\eea}{\end{eqnarray}}
\newcommand{\eeaa}{\end{eqnarray*}}

\begin{document}

\title{\bf\Large {Model for the FC and ZFC Ferrimagnetic Spinel}}

\author{N. Karchev}
\affiliation{Department of Physics, Sofia University, James Bourchier 5 blvd., 1164
Sofia, Bulgaria}

\date{\today}

\begin{abstract}
There are two methods of preparation of ferrimagnetic spinel. If, during the preparation, an external magnetic field as high as 300 O\"{e} is applied upon cooling the material is named field-cooled (FC). If the applied field is about 1O\"{e} the material is zero-field cooled (ZFC). To explore the magnetic and thermodynamic  properties of these materials we consider two-sublattice spin system, defined on the bcc lattice, with spin-$s^A$ operators $\bf{S_{i}^A}$ at the sublattice $A$ site and spin-$s^B$ operators $\bf{S_{i}^B}$ at the sublattice $B$ site, where $s^A>s^B$. The subtle point is the exchange between sublattice A and B spins, which is antiferromanetic. Applying  magnetic field along the sublattice A magnetization, during preparation of the material,  one  compensates the Zeeman splitting, due to the exchange, of sublattice B electrons. This effectively leads to a decrease of the $s^B$ spin. We consider a model with $s^B$ varying parameter which accounts for the applied, during the preparation, magnetic field.

It is shown that the model agrees well with the observed magnetization-temperature curves of zero field cooled (ZFC) and non-zero field cooled (FC) spinel ferrimagnetic spinel and explains the anomalous temperature dependence of the specific heat.

\end{abstract}

 \maketitle

\section{Introduction}

The magnetization-temperature and magnetic susceptibility curves for zero field cooled (ZFC) and non-zero field cooled (FC) ferrimagnetic spinel display a notable difference below N\'{e}el $T_N$ temperature \cite{spinel+,spinelFeCr2S4,spinelCv1,spinel08,spinel++,spinel11b,spinel11a,spinel11c,spinel12a,spinel12b,spinel+1,spinel+2,spinel+3}. The (ZFC) curve exhibits a maximum and then a monotonic decrease upon cooling from $T_N$, while the (FC) curve increases steeply, shows a dip near the temperature at which the (ZFC) curve has a maximum and finally increases monotonically. There is also difference between so-called field-cooled-cooling and field-cooled-warming procedures \cite{FCC-FCW}.

The specific heat curves for spinel ferrimagnetics   show sharp peak at N\'{e}el temperature, which indicates ferrimagnetic to paramagnetic transition, and tiny peak at temperature below N\'{e}el's where the magnetization-temperature curve has a maximum \cite{spinelCv1,spinel11b,spinel12a}.

Although the FC and ZFC spinel ferrimagnetics have intensively been studied, their magnetic and thermodynamic properties have not been understood. There is not an effective model which in unified way to explain the experimental results.

In the present paper we consider two-sublattice spin system, defined on the bcc lattice, with spin-$s^A$ operators $\bf{S_{i}^A}$ at the sublattice $A$ site and spin-$s^B$ operators $\bf{S_{i}^B}$ at the sublattice $B$ site, where $s^A>s^B$. The subtle point is the exchange between sublattice A and B spins, which is antiferromanetic. Applying  magnetic field along the sublattice A magnetization, during preparation of the material,  one  compensates the Zeeman splitting of sublattice B electrons due to the exchange. This effectively leads to decrease of the $s^B$ spin. One can obtain an intuition for this from spin-fermion model of spinel ferrimagnetic  with spin-$1/2$ itinerant electrons at the sublattice $B$ site and spin-$s$ localized electrons  at the sublattice $A$ site. An applied, along the magnetization of the localized electrons, external magnetic field compensates the Zeeman splitting due to the spin-fermion exchange. Integrating out the fermions we obtain an effective spin model with effective spin $s^B$ which depends on the external magnetic field (see Appendix A).

We consider a model with $s^B$ varying parameter which accounts for the applied, during the preparation, magnetic field. First we present the method of calculation exploring ZFC spinel with $s^A=1.5$ and $s^B=1$. We obtained that the system has two phases. At low temperature $(0,T^*)$ the magnetic orders of the $A$ and $B$ spins  contribute to the magnetization of the system, while at the high temperature  $(T^*,T_N)$, the magnetic order of the sublattice B with smaller spin $s^B$ and with a weaker intra-sublattice exchange is suppressed by magnon fluctuations. Only the sublattice A spins, with stronger intra-sublattice exchange, have non-zero spontaneous magnetization. There is no additional symmetry breaking, and the Goldstone boson has a ferromagnetic dispersion in both phases, but partial-order transition demonstrates itself through tiny peak of specific heat as a function of temperature.

Partial order is well known phenomenon and has been subject to extensive studies. Frustrated antiferromagnetic systems has been studied by means of Green function formalism. Partial order and anomalous temperature dependence of specific heat have been predicted \cite{Diep97}. Experimentally the partial order  has been observed in $Gd_2Ti_2O_7$ \cite{POexp04}. Monte Carlo method has been utilized to study the nature of partial order in Ising model on  $kagom\acute{e}$ lattice \cite{Diep87}. There are exact results for the partially ordered systems which precede the above studies \cite{Larkin66,Diep87,Diep04}. The advantage of the present method of calculation is that it permits to consider the sublattice B spin as a varying parameter $s^B<1$. We calculate the magnetization-temperature curves and the specific heat as a function of temperature for $s^B=0.7$, $s^B=0.4$, $s^B=0.2$ and fixed $s^A=1.5$,  thus accounting for the increasing of the applied, during preparation of FC spinel, magnetic field.

\section{Method of calculation}

The Hamiltonian of the ZFC ferrimagnetic spinel is
\bea \label{fc-ferri1}
 H & = & - J^A\sum\limits_{\ll ij \gg _A } {{\bf S}^A_{i}
\cdot {\bf S}^A_{j}}\,-\,J^B\sum\limits_{\ll ij \gg _B } {{\bf S}^B_{i}
\cdot {\bf S}^B_{j}}\nonumber \\
& & +\,J \sum\limits_{\langle ij \rangle} {{\bf S}^A_{i}
\cdot {\bf S}^B_{j}}
  \eea where the sums are over all sites of a body-centered cubic (bcc) lattice:
$\langle i,j\rangle$ denotes the sum over the nearest neighbors, $\ll i,j \gg _A$ denotes the sum over the sites of the A sublattice, $\ll i,j \gg _B$ denotes the sum over the sites of the B sublattice.
The first two terms  describe the ferromagnetic Heisenberg intra-sublattice
exchange $J^A>0, J^B>0$, while the third term describes the inter-sublattice exchange which is antiferromagnetic $J>0$.

To study a theory with the Hamiltonian Eq.(\ref{fc-ferri1}) it is convenient to introduce Holstein-Primakoff representation for the spin
operators ${\bf S}^A_{i}(a^+,a)$ and ${\bf S}^B_{i}(b^+,b)$. Rewriting the effective Hamiltonian in terms of the Bose operators $(a^+,a,b^+,b)$ we keep  only the quadratic and quartic terms. The next step is to represent the Hamiltonian in the Hartree-Fock  approximation:
\be\label{fc-ferri2}H\approx H_{HF}=H_{cl}+H_q\ee
with
\bea\label{fc-ferri3} H_{cl} & = & 6 N J^A (s^A)^2 (u^A-1)^2+ 6 N J^B (s^B)^2 (u^B-1)^2 \nonumber \\
& + & 8 N J s^A s^B (u-1)^2,\eea
and
\be\label{fc-ferri4}
H_q = \sum\limits_{k\in B_r}\left [\varepsilon^a_k\,a_k^+a_k\,+\,\varepsilon^b_k\,b_k^+b_k\,-
\,\gamma_k \left (a_k^+b_k^+ + b_k a_k \right )\,\right ],
\ee
where $N=N^A=N^B$ is the number of sites on a sublattice. The two equivalent sublattices A and B of the bcc lattice are simple cubic lattices. The wave vector $k$ runs over the reduced  first Brillouin zone $B_r$ of a
bcc lattice which is the first Brillouin zone of a simple cubic lattice. The dispersions are given by equalities
\bea\label{fc-ferri5}
\varepsilon^a_k & = & 4s^A J^A u^A \left(3-\cos k_x-\cos k_y - \cos k_z\right)
\,+\,8s^B\,J u \nonumber\\
\varepsilon^b_k & = & 4s^B J^B u^B \left(3-\cos k_x-\cos k_y - \cos k_z\right)
\,+\,8s^A\,J u \nonumber\\
\gamma_k & = & 8J\,u\,\sqrt{s^A\,s^B}\,\cos \frac {k_x}{2} \,\cos \frac {k_y}{2} \,\cos \frac {k_z}{2} \eea
The equations (\ref{fc-ferri5}) show that Hartree-Fock parameters ($u^A,u^B,u$) renormalize the intra and inter-sublattice exchange constants
 ($J^A, J^B,J$) respectively.

 To diagonalize the Hamiltonian one introduces new Bose fields
$\alpha_k,\,\alpha_k^+,\,\beta_k,\,\beta_k^+$ by means of the
transformation
\bea \label{fc-ferri6} & &
a_k\,=u_k\,\alpha_k\,+\,v_k\,\beta^+_k\qquad
a_k^+\,=u_k\,\alpha_k^+\,+\,v_k\,\beta_k
\nonumber \\
\\
& & b_k\,=\,u_k\,\beta_k\,+\,v_k\,\alpha^+_k\qquad
b_k^+\,=\,u_k\,\beta_k^+\,+\,v_k\,\alpha_k,
\nonumber \eea where the coefficients of the transformation $u_k$ and $v_k$ are real functions of the wave vector $k$ (\ref{ferri11}).
The transformed Hamiltonian adopts the form \be
\label{fc-ferri7} H_q = \sum\limits_{k\in B_r}\left
(E^{\alpha}_k\,\alpha_k^+\alpha_k\,+\,E^{\beta}_k\,\beta_k^+\beta_k\,+\,E^0_k\right),
\ee
with new dispersions \bea  \label{fc-ferri8} & & E^{\alpha}_k\,=\,\frac
12\,\left [
\sqrt{(\varepsilon^a_k\,+\,\varepsilon^b_k)^2\,-\,4\gamma^2_k}\,-\,\varepsilon^b_k\,+\,\varepsilon^a_k\right] \nonumber \\
\\
& & E^{\beta}_k\,=\,\frac
12\,\left [
\sqrt{(\varepsilon^a_k\,+\,\varepsilon^b_k)^2\,-\,4\gamma^2_k}\,+\,\varepsilon^b_k\,-\,\varepsilon^a_k\right]
\nonumber \eea and vacuum energy
\be\label{fc-ferri9}
 E^{0}_k\,=\,\frac
12\,\left [
\sqrt{(\varepsilon^a_k\,+\,\varepsilon^b_k)^2\,-\,4\gamma^2_k}\,-\,\varepsilon^b_k\,-\,\varepsilon^a_k\right]\ee
For positive values of the Hartree-Fock parameters and all values of $k\in B_r$,\,
the dispersions are nonnegative $ E^{\alpha}_k\geq 0,\, E^{\beta}_k \geq 0$. When $s^A>s^B$
the $\alpha_k$ boson is the long-range \textbf{(magnon)} excitation in the system with $E^{\alpha}_k\propto\rho k^2$, near the zero wavevector, while the $\beta_k$ boson is a gapped excitation, with gap proportional to the inter-sublattice exchange constant $E^{\beta}_0=8Ju(s^A-s^B)$  .

The free energy of a system with Hamiltonian $H_{HF}$ equations (\ref{fc-ferri2}), (\ref{fc-ferri3}) and  (\ref{fc-ferri4}) is
\bea\label{fc-ferri10}
\mathcal{F} & = & 6 N J^A (s^A)^2 (u^A-1)^2+ 6 N J^B (s^B)^2 (u^B-1)^2 \nonumber \\
& + & 8 N J s^A s^B (u-1)^2 + \frac 1N \sum\limits_{k\in B_r}E^{0}_k \\
& + & \frac {1}{\beta N} \sum\limits_{k\in B_r}\left[ \ln\left(1-e^{-\beta E^{\alpha}_k}\right)\,+\,\ln\left(1-e^{-\beta E^{\beta}_k}\right)\right],\nonumber\eea
where $\beta\,=\,1/T$\,\, is the inverse temperature.
Then, the system of equations for the Hartree-Fock parameters is
\be\label{fc-ferri11}\partial\mathcal{F}/\partial u^A=0,\quad \partial\mathcal{F}/\partial u^B=0,\quad\partial\mathcal{F}/\partial u=0.\ee

The Hartree-Fock parameters are positive functions of $T/J$ , solution of the system of equations (\ref{fc-ferri11}) (see Eqs.(\ref{fc-ferri11a})). Utilizing these functions, one can calculate the spontaneous magnetization on the two sublattices
\bea \label{fc-ferri12}
M^A & = & <S^3_{1j}> \,\,\, j\,\, is\,\, from\,\, sublattice\,\, A \nonumber \\
\\
M^B & = & <S^3_{2j}> \,\,\, j\,\, is\,\, from\,\, sublattice\,\, B \nonumber \eea
and $M\,=\,M^A\,+\,M^B$, the spontaneous magnetization of the system.
In terms of the Bose functions of the $\alpha$ and $\beta$ excitations they adopt the form
\bea\label{fc-ferri13}
M^A & = & s^A\,-\,\frac 1N \sum\limits_{k\in B_r} \left[u_k^2 \,n_k^{\alpha}\, +\, v_k^2\, n_k^{\beta}\, +\, v_k^2\right] \\
M^B & = & - \,s^B\,+\,\frac 1N \sum\limits_{k\in B_r} \left[v_k^2 \,n_k^{\alpha}\, +\, u_k^2\, n_k^{\beta}\, +\, v_k^2\right],\nonumber \eea
where $u_k$ and $v_k$ are functions of the wavevector $k$, coefficients in the transformation (\ref{fc-ferri6}).

\begin{figure}[tb]
\centerline{\epsfig{width=75mm,file=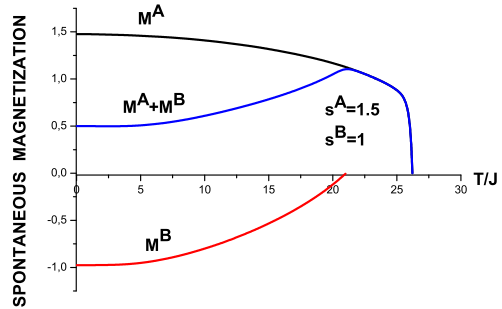}}
\caption{(Color online) Spontaneous magnetization $M^A$, $M^B$ and $M^A+M^B$ as a function of the temperature in units of exchange constant $T/J$ for ferrimagnetic spinel on bcc lattice with $s^A=1.5$, $s^B=1$, $J^A/J=2$ and $J^B/J=0.002$. The partial order transition temperature $T^*$ is the temperature above which $M^B=0$.}
\label{zfc-ferri-MMM}
\end{figure}

The magnon excitation - $\alpha_k$ is a complicated mixture of the transversal
fluctuations of the $A$ and $B$ spins (\ref{fc-ferri6}). As a result the magnons' fluctuations suppress in a different way the magnetization on sublattices $A$ and $B$. Quantitatively this depends on the coefficients $u_k$ and $v_k$. At characteristic temperature $T^*$  spontaneous magnetization on sublattice $B$ becomes equal to zero, while spontaneous magnetization on sublattice $A$ is still nonzero (see Fig.\ref{zfc-ferri-MMM}).

To study the magnetic properties of the system above $T^*$ we first consider the paramagnetic phase. To this end we make use of the Takahashi modified spin-wave theory \cite{Takahashi87} and introduce two parameters $\lambda^A$ and $\lambda^B$  to enforce the magnetization on the two sublattices to be equal to zero. The new Hamiltonian is obtained from the old one equation (\ref{fc-ferri1}) by adding two new terms:
\be
\label{fc-ferri14} \hat{H}\,=\,H\,-\,\sum\limits_{i\in A}
\lambda^A S^{A3}_{i}\,+\,\sum\limits_{i\in B} \lambda^B S^{B3}_{i} \ee
In momentum space the new Hamiltonian $\hat{H}$ adopts the form Eq.(\ref{fc-ferri4}) with new dispersions
 \be \label{ferri28}
\hat{\varepsilon}^a_k\,=\varepsilon^a_k\,+\,\lambda^A, \qquad
\hat{\varepsilon}^b_k\,=\varepsilon^b_k\,+\,\lambda^B.\ee
Utilizing the same transformation (\ref{fc-ferri6}) one obtains the Hamiltonian $\hat{H}$ in diagonal form (\ref{fc-ferri7}) with dispersions $\hat{E^{\alpha}}$ and $\hat{E^{\beta}}$ obtained from Eqs.(\ref{fc-ferri8}) replacing $\varepsilon^a_k$ and $\varepsilon^b_k$ with $\hat{\varepsilon}^a_k$ and $\hat{\varepsilon}^b_k$ respectively.
It is convenient to represent the parameters
$\lambda^A$ and $\lambda^B$ in the form
\be \label{fc-ferri15}
\lambda^A\,=\,6 J u s^B (\mu^A\,-\,1),\quad
\lambda^B\,=\,6 J u s^A (\mu^B\,-\,1). \ee
The dispersions $\hat{\varepsilon}^a_k$ and $\hat{\varepsilon}^b_k$
are positive for all values of the wavevector $k$, if the parameters  $\mu^A$ and $\mu^B$ are positive.
The dispersions $\hat{E^{\alpha}_k}$ and $\hat{E^{\beta}_k}$ are well defined if
\be\label{fc-ferri16}
\mu^A\mu^B\geq1.\ee
The $\beta_k$ excitation is gapped ($\hat{E}^{\beta}_k>0 $)
for all values of parameters $\mu^A$ and $\mu^B$ which satisfy equation (\ref{fc-ferri16}). The $\alpha$ excitation is gapped if $\mu^A
\mu^B>1$, but in the particular case
\be \label{fc-ferri17}
\mu^A\mu^B=1\ee
$\hat{E}^{\alpha}_0=0$, and near the zero wavevector $\hat{E}^{\alpha}_k\approx \hat{\rho} k^2$. Therefor, in the particular case Eq. (\ref{fc-ferri17}) $\alpha_k$ boson is the long-range excitation (magnon) in the system.

Above N\'{e}el temperature we introduced the parameters $\mu^A$ and  $\mu^B$($\lambda^A,\lambda^B$) to enforce the sublattice $A$ and $B$ spontaneous magnetizations to be equal to zero. We find out these and Hartree-Fock parameters, as functions of temperature, solving the system of five equations, equations (\ref{fc-ferri11}) and the equations $M^A=M^B=0$, where the  spontaneous magnetization has the same representation as equations (\ref{fc-ferri13}) but with coefficients $\hat{u}_k,\,\,
\hat{v}_k$, and dispersions $\hat{E}^{\alpha}_k,\,\, \hat{E}^{\beta}_k$ in the expressions for the Bose functions.
The numerical calculations show that above N\'{e}el temperature
$\mu^A\mu^B>1$. When the temperature decreases the product $\mu^A\mu^B$ decreases, remaining larger than one. The temperature at which the product becomes equal to one ($\mu^A\mu^B=1$) is the N\'{e}el temperature Fig.(\ref{mumumu}).

Below $T_N$, the spectrum contains long-range (magnon) excitations. Thereupon, $\mu^A\mu^B=1$ and one can use the representation
$\mu^B=1/\mu^A$. Then, $\mu^A$ and  Hartree-Fock parameters are solution of a system of four equations, equations (\ref{fc-ferri11}) and the equation $M^B=0$.

We utilize the
obtained functions $\mu^A(T)$,$\mu^B(T)$ $u^A(T)$, $u^B(T)$, $u(T)$ (see Appendix C) to calculate the spontaneous magnetization as a function of the temperature. For a system with $s^A=1.5$, $s^B=1$, $J^A/J=2$ and $J^B/J=0.002$ the functions $M^A(T/J)$, $M^B(T/J)$ and $M^A(T/J)+M^B(T/J)$ are depicted in figure (\ref{zfc-ferri-MMM}).
The partial order transition temperature $T^*$ is the temperature above which $M^B=0$.

The customary formula for the entropy of a Bose system with Hamiltonian (\ref{fc-ferri7}) is
\be\label{fc-ferri18}
\mathcal{S}=\frac 1N \sum\limits_{k,\delta}\left [(1+n^{\delta}_k)\ln (1+n^{\delta}_k)-n^{\delta}_k\ln n^{\delta}_k \right ], \ee
where $\delta$ stays for $\alpha$ and $\beta $. The dispersions $E^{\alpha}_k$ and $E^{\beta}_k$ (\ref{fc-ferri8}) are used to define the Bose functions $n^{\alpha}_k$ and $n^{\beta}_k$ below $T^*$, dispersions $\hat {E^{\alpha}_k}$ and  $\hat {E^{\beta}_k}$ with  $\mu^B=1/\mu^A$ are used for partial order phase $T^*<T<T_N$, and with  $\mu^A\mu^B>1$ for paramagnetic phase above  N\'{e}el temperature.
With entropy, as a function of temperature in mind, one can calculate the contribution of magnons to the specific heat:
\be\label{QCB50}
C=T \frac {d \mathcal{S}}{dT} \ee
The resultant curve $C(T/J)$, for a system with the same parameters, as above, is depicted in figure (\ref{fc-ferri-Cv}).
\begin{figure}[tb]
\centerline{\epsfig{width=75mm,file=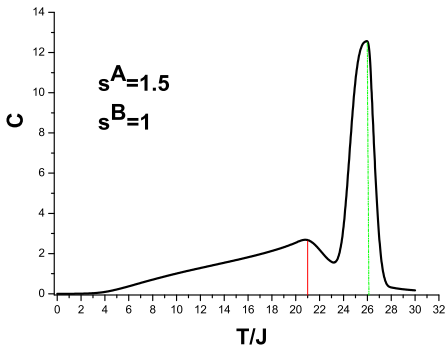}}
\caption{(Color online) Specific heat $C$ vs temperature, in units of  the exchange constant $T/J$,
 for $J^A/J=2$, $J^B/J=0.002$, $s^A=1.5$ and $s^B=1$. The high temperature (green) vertical line marks N\'{e}el $T_N$ temperature, while the low temperature (red) line marks partial order transition temperature $T^*$.}
\label{zfc-ferri-Cv}
\end{figure}

The figures (\ref{zfc-ferri-MMM}) and (\ref{zfc-ferri-Cv}) show that the present method of calculation describes correctly the features of the system, partial order transition and anomalous behavior of the specific heat at the temperature of this transition $T^*$.

\section{FC spinel ferrimagnetics}

FC spinel ferrimagnetics are prepared applying magnetic field  along the sublattice A magnetization upon cooling the material. This compensates the Zeeman splitting of sublattice B electrons which effectively leads to decrease of the $s^B$ spin. The Hamiltonian of the FC spinel ferrimagnet  is
given by Eq.(\ref{fc-ferri1}) with $s^B$ a varying parameter. The advantage of the method of calculation, presented above, is that we can use the same systems of equations for the three phases, $0<T<T^*$, $T^*<T<T_N$, $T>T_N$, and different values of the parameter $s^B$. We calculate the magnetization-temperature curves and the specific heat as a function of temperature for $s^A=1.5$ and three different values of $s^B$ ($s^B=0.7$, $s^B=0.4$ and $s^B=0.2$), thus accounting for the increasing of the applied, during preparation of FC spinel, magnetic field.

\begin{figure}[tb]
\centerline{\epsfig{width=70mm,file=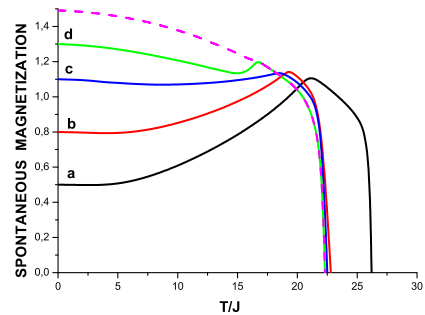}}
\caption{Color online) Spontaneous magnetization $M^A+M^B$ as a function of the temperature in units of the exchange constant $T/J$ for ferrimagnetic spinel on bcc lattice with $J^A/J=2$, $J^B=0.002$, $s^A=1.5$ and :a)(black) $s^B=1$, b)(red) $s^B=0.7$, c)(blue) $s^B=0.4$ and d)(green) $s^B=0.2$. The dash (magenta) line shows the spontaneous magnetization $M^A$ of sublattice $A$ for $s^A=1.5$ and $s^B=0.2$.}
\label{fc-ferri-M}
\end{figure}

The resultant magnetization-temperature curves are depicted in figure (\ref{fc-ferri-M}).
The curve "a" corresponds to ZFC spinel (see figure (\ref{fc-ferri1})). The increasing of the applied, during the preparation, magnetic field is modeled by decreasing of $s^B$. The curves "b", "c" and "d" are magnetization-temperature curves for FC spinel ferrimagnetics with  $s^B=0.7$, $s^B=0.4$ and $s^B=0.2$. The curve "c" agrees well with the observed magnetization-temperature curve of FC spinel  $MnV_2O_4$ with an external magnetic field as high as 300 O\"{e} applied upon cooling the material \cite{spinel08,spinel++}. The dash line shows the spontaneous magnetization $M^A$ of sublattice $A$ for $s^A=1.5$ and $s^B=0.2$. Comparing with line "d", which shows the spontaneous magnetization $M^A+M^B$ of the same system one can conclude that contribution of the sublattice B magnetization is small and Zeeman splitting, of sublattice B electrons, is approximately compensated.

\begin{figure}[tb]
\centerline{\epsfig{width=70mm,file=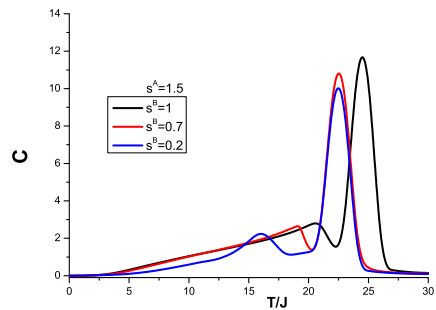}}
\caption{(Color online) Specific heat $C$ vs temperature, in units of  the exchange constant $T/J$, for $s^A=1.5$ and different values of the effective spin $s^B$: black curve $s^B=1$, red curve $s^B=0.7$ and blue curve $s^B=0.2$ .}
\label{fc-ferri-Cv}
\end{figure}

The dependence of specific heat on temperature for different values of $s^B$ is shown in figure (\ref{fc-ferri-Cv}).
The curve with higher maximum corresponds to ZFC spinel (see figure (\ref{zfc-ferri-Cv}))
With increasing the applied, during the preparation, magnetic field (with decreasing $s^B$) the anomalous temperature behavior of specific heat, at temperature of partial order transition, remains well defined but the pick decreases. The anomalous pick is well observed experimentally \cite{spinelCv1,spinel11b,spinel12a}, and one can use it to determine the $T^*$ temperature of the partial order transition.

\section{\bf Summary}

In this paper is pointed out that a model which agrees well with the observed magnetic and thermodynamic properties of ZFC and FC spinel ferrimagnets is a two-sublattice Heisenberg model with spin-$s^A$ operators at the sublattice $A$ site, spin-$s^B$ operators at the sublattice $B$ site, ferromagnetic intra-sublattice exchange and antiferromagnetic inter-sublattice exchange. The applied magnetic field, along the sublattice A magnetization upon cooling the material, is accounted for varying only one parameter $s^B$.

The subtle point is that $MnV_2O_4$ spinel has an obvious anomalous  magnetic behavior, but the appearance of this anomaly is attributed to the strong orbital-spin coupling \cite{spinel08} which is not discussed in the model (\ref{fc-ferri1}). The spinel $MnV_2O_4$ is a two-sublattice ferrimagnet, with site $A$ occupied by the $Mn^{2+}$ ion, which is in the $3d^5$ high-spin configuration with quenched orbital angular momentum, which can be regarded as a  $s=5/2$ spin. The B site is occupied by the $V^{3+}$ ion, which takes the $3d^{2}$ high-spin configuration in the triply degenerate $t_{2g}$ orbital and has orbital degrees of freedom. Because of the strong spin-orbital interaction it is convenient to consider $jj$ coupling with $\textbf{J}^A=\textbf{S}^A$ and $\textbf{J}^B=\textbf{L}^B+\textbf{S}^B$. The sublattice $A$ total angular momentum is $j^A=s^A=5/2$, while the sublattice $B$ total angular momentum is $j^B=l^B+s^B$, with $l^B=3$, and $s^B=1$ \cite{spinel+}.
Then the g-factor for the sublattice $A$ is $g^A=2$, and for the sublattice $B$  $g^B=\frac 54$. The sublattice $A$ magnetic order is antiparallel to the sublattice $B$ one and the saturated magnetization is $\sigma=2 \frac 52-\frac 54 4=0$, in agreement with the experimental finding for ZFC spinel that the magnetization goes to zero when the temperature approaches zero.
The Hamiltonian of the system is
\bea \label{MnV1}
 H & = & - \kappa^A\sum\limits_{\ll ij \gg _A } {{\bf J}^A_{i}
\cdot {\bf J}^A_{j}}\,-\,\kappa^B\sum\limits_{\ll ij \gg _B } {{\bf J}^B_{i}
\cdot {\bf J}^B_{j}}\nonumber \\
& & +\,\kappa \sum\limits_{\langle ij \rangle} {{\bf J}^A_{i}
\cdot {\bf J}^B_{j}}\eea
The first two terms  describe the ferromagnetic Heisenberg intra-sublattice
exchange $\kappa^A>0, \kappa^B>0$, while the third term describes the inter-sublattice exchange which is antiferromagnetic $\kappa>0$.
The operators ${\bf J}^A_{j}$ and ${\bf J}^B_{j}$ satisfy the $SU(2)$ algebra, therefor we can use the Holstein-Primakoff representation of the total angular momentum vectors ${\bf J}^A_{j}(a^+_j,a_j)$ and ${\bf J}^B_{j}(b^+_j,\,b_j)$, where $a^+_j,\,a_j$
and $b^+_j,\,b_j$ are Bose fields. Farther on, we repeat the calculations from sections II and III. The only difference is the expression for the magnetization $g^A\,M^A\,+\,g^B\,M^B$. The $jj$ model shows that the anomalous magnetic and thermodynamic behavior of $MnV_2O_4$ is a consequence of the ferrimagnetic nature of the system.

\appendix
\section{Spin-fermion model of spinel ferrimagnet}

The Hamiltonian of the spin-fermion model of ferrimagnetic spinel defined on a body centered cubic lattice is
\bea \label{IFerri1}\nonumber
H  = & - & t\sum\limits_{\ll ij \gg _B } {\left( {c_{i\sigma }^ + c_{j\sigma } + h.c.} \right)}
  -\mu \sum\limits_{i\in B} {n_i
} \\ & - & J^{B1}\sum\limits_{  \ll  ij  \gg_B  } {{\bf S_i^B}
\cdot {\bf S_j^B}}
  + J\sum\limits_{  \langle  ij  \rangle } {{\bf
S_i^A}}\cdot {\bf S_j^B}  \\
& - & J^A\sum\limits_{  \ll  ij  \gg_A  } {{\bf S_i^A}
\cdot {\bf S_j^A}}
- H \sum\limits_{i\in A} {S^{zA}_{i}}-H \sum\limits_{i\in B} {S^{zB}_i}, \nonumber
\eea
where $S^{\nu B}_i=\frac 12\sum\limits_{\sigma\sigma'}c^+_{i\sigma}\tau^{\nu}_{\sigma\sigma'}c^{\phantom +}_{i\sigma'}$, with the Pauli
matrices $(\tau^x,\tau^y,\tau^z)$, is the spin of the itinerant
electrons at the sublattice $B$ site , ${\bf S_i^A}$ is the spin of the localized electrons  at the sublattice $A$ site, $\mu$
is the chemical potential, and $n_i=c^+_{i\sigma}c_{i\sigma}$. The
sums are over all sites of a body centered cubic lattice, $\langle i,j\rangle$ denotes the sum over the nearest neighbors, while $ \ll  ij  \gg_A$ and  $\ll  ij  \gg_B$ are sums over all sites of sublattice $A$ and $B$ respectively. The Heisenberg term $(J^A > 0)$ describes ferromagnetic Heisenberg
exchange between localized electrons and $J>0$ is the antiferromagnetic exchange constant between localized and itinerant electrons. $H>0$ is the Zeeman splitting energy due to the external magnetic field (magnetic field in units of energy). We represent the Fermi operators, the spin of the itinerant electrons and the density operators $n_{i\sigma}$  in terms of the Schwinger bosons
($\varphi_{i,\sigma}, \varphi_{i,\sigma}^+$) and slave fermions
($h_i, h_i^+,d_i,d_i^+$). The Bose fields
are doublets $(\sigma=1,2)$ without charge, while fermions
are spinless with charges 1 ($d_i$) and -1 ($h_i$):
\begin{eqnarray}\label{QCB2} & & c_{i\uparrow} =
h_i^+\varphi _{i1}+ \varphi_{i2}^+ d_i, \qquad c_{i\downarrow} =
h_i^+ \varphi _{i2}- \varphi_{i1}^+ d_i, \nonumber
\\
& & n_i = 1 - h^+_i h_i +  d^+_i d_i,\quad  s^{\nu}_i=\frac 12
\sum\limits_{\sigma\sigma'} \varphi^+_{i\sigma}
{\tau}^{\nu}_{\sigma\sigma'} \varphi_{i\sigma'},\nonumber
\\& &
c_{i\uparrow }^+c_{i\uparrow }c_{i\downarrow }^+c_{i\downarrow}=d_i^+d_i \eea

\be\label{QCB2b}
\varphi_{i1}^+ \varphi_{i1}+ \varphi_{i2}^+ \varphi_{i2}+ d_i^+
d_i+h_i^+ h_i=1  \ee
To solve the constraint (Eq.\ref{QCB2b}), one makes a change of variables, introducing
Bose doublets $\zeta_{i\sigma}$ and
$\zeta^+_{i\sigma}\,$\cite{Schmeltzer}
\begin{eqnarray}\label{QCB3}
\zeta_{i\sigma} & = & \varphi_{i\sigma} \left(1-h^+_i h_i-d^+_i
d_i\right)^
{-\frac 12},\nonumber \\
\zeta^+_{i\sigma} & = & \varphi^+_{i\sigma} \left(1-h^+_i h_i-d^+_i
d_i\right)^ {-\frac 12},
\end{eqnarray}
where the new fields satisfy the constraint
$\zeta^+_{i\sigma}\zeta_{i\sigma}\,=\,1$. In terms of the new fields
the spin vectors of the itinerant electrons have the form
\be
S^{\nu B}_i=\frac 12 \sum\limits_{\sigma\sigma'} \zeta^+_{i\sigma}
{\tau}^{\nu}_{\sigma\sigma'} \zeta_{i\sigma'} \left[1-h^+_i
h_i-d^+_i d_i\right] \label{QCB4} \ee
When, in the ground state,
the lattice site is empty, the operator identity $h^+_ih_i=1$ is
true. When the lattice site is doubly occupied, $d^+_id_i=1$. Hence,
when the lattice site is empty or doubly occupied the spin on this
site is zero. When the lattice site is neither empty nor doubly
occupied ($h^+_ih_i=d^+_id_i=0$), the spin equals $\,\,{\bf s}_{i}=1/2
{\bf n}_i,\,\,$ where the unit vector
\be\label{QCB5b}
n^{\nu}_i=\sum\limits_{\sigma\sigma'} \zeta^+_{i\sigma}
{\tau}^{\nu}_{\sigma\sigma'} \zeta_{i\sigma'}\qquad ({\bf
n}_i^2=1)\ee identifies the local orientation of the spin of the
itinerant electron.

The part of Hamiltonian Eq.(\ref{IFerri1}) with itinerant fermions can be rewritten in terms of Bose fields Eq.(\ref{QCB3}) and slave fermions
\bea\label{QCB3a}
H^B  & = & -t\sum\limits_{{\ll ij \gg _B }} \left[\left ( d^+_j d_i-h^+_j h_i \right) \zeta^+_{i\sigma}\zeta_{j\sigma}\right. \nonumber \\
& + & \left.\left ( d^+_j h^+_i-d^+_i h^+_j\right )\left (\zeta_{i1}\zeta_{j2}-\zeta_{i2}\zeta_{j1}\right ) + h.c. \right]\nonumber \\
& \times & \left(1-h^+_i h_i-d^+_id_i\right)^{\frac 12}\left(1-h^+_j h_j-d^+_jd_j\right)^{\frac 12} \nonumber \\
&-& J^{B1} \sum\limits_{{\ll ij \gg _B }} m_i m_j\, {{\bf n_i}}\cdot {\bf n_j} \nonumber \\
& + & U \sum\limits_{i\in B} d^+_id_i -\mu \sum\limits_{i\in B} \left (1-h^+_ih_i+d^+_id_i\right)\\
& - & H \sum\limits_{i\in B}\frac 12 n_i^z[1-h^+_ih_i- d^+_id_i]. \nonumber\eea
where
\be\label{QCB3c}
m_i=\frac 12 [1-h_i^+h_i-d_i^+d_i].\ee
The mixed, spin-fermion term, adopts the form
\be\label{QCB3b}
H^{AB}= J\sum\limits_{  \langle  ij  \rangle } {{\bf S_i^A}}\cdot {\bf n_j}\, m_j \ee

The Hamiltonian (\ref{IFerri1}) can be rewritten in the form
\be H=H^A+H^B+H^{AB}\ee
where $H^A$ is the contribution of sublattice A spins ${\bf S_i^A}$.

An important advantage of working with Schwinger bosons and slave fermions
is the fact that Hubbard term is in a diagonal form. The fermion-fermion and fermion-boson interactions are included in the hopping term. One treats them as a perturbation. To proceed we approximate the hopping term of the Hamiltonian Eq.(\ref{QCB3a}) setting  $\left(1-h^+_i h_i-d^+_id_i\right)^{\frac 12}\sim 1$ and keeping only the quadratic, with respect to fermions, terms. This means that the averaging in the subspace of the fermions is performed in one fermion-loop approximation. Further, we represent the resulting $h^B$ and $h^{AB}$ Hamiltonian as a sum of two terms
\be\label{QCB4a}
H^f=H^f_0 + H^{fb}_{int}, \ee
where
\bea\label{QCB4b}
H^f_0 = & - & t\sum\limits_{{\ll ij \gg _B }} \left ( d^+_j d_i-h^+_j h_i + h.c.\right)
 +  U \sum\limits_{i\in B} d^+_id_i \nonumber \\
& - & \mu \sum\limits_{i\in B} \left (1-h^+_ih_i+d^+_id_i\right)\\
& + & \frac {4sJ-H}{2}\sum\limits_{i\in B} \left (h^+_ih_i+d^+_id_i\right)\nonumber ,\eea
is the Hamiltonian of the free $d$ and $h$ fermions, and
\bea\label{QCB4c}
H^{fb}_{int}  = & - & t\sum\limits_{{\ll ij \gg _B }} \left[\left ( d^+_j d_i-h^+_j h_i \right) \left (\zeta^+_{i\sigma}\zeta_{j\sigma}-1\right)\right. \\
& + & \left.\left ( d^+_j h^+_i-d^+_i h^+_j\right )\left (\zeta_{i1}\zeta_{j2}-\zeta_{i2}\zeta_{j1}\right ) + h.c. \right]\nonumber \\
&-& J^{B1} \sum\limits_{{\ll ij \gg _B }}m_i m_j\, {{\bf n_i}}\cdot {\bf n_j} \nonumber \eea
is the Hamiltonian of boson-fermion interaction.

The ground state of the system, without accounting for the spin fluctuations, is determined by the free-fermion Hamiltonian $h_0$ and is labeled by the density of electrons
\be\label{QCB4d} n=1-<h^+_i h_i>+<d^+_id_i> \ee (see equation (\ref{QCB2})) and the "effective spin" of the sublattice B electron
\begin{equation}
s^B=\frac 12 \left(1-<h^+_i h_i>-<d^+_id_i>\right). \label{QCB5}
\end{equation}
At half-filling
\be\label{QCB4e} <h^+_i h_i>=<d^+_id_i>. \ee To solve this equation one sets the chemical potential $\mu=U/2$. Utilizing this representation of $\mu$ we calculate the effective spin $s^B$ as a function of applied magnetic field $h$ for parameters $4t/U=2.2$ and $4s^AJ/U=4$.
The result is depicted in figure (\ref{fc-ferri-spin}).
\begin{figure}[tb]
\centerline{\epsfig{width=85mm,file=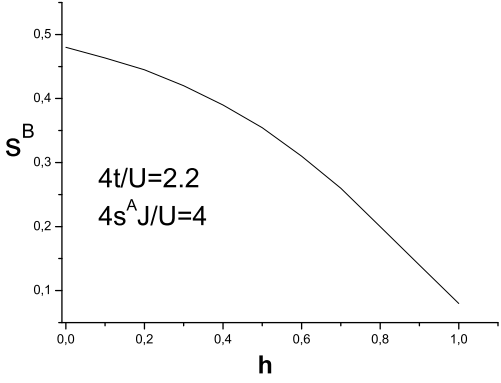}}
\caption{(Color online) Sublattice B effective spin $s^B$ as a function of applied magnetic field $h$ for parameters $4t/U=2.2$ and $4s^AJ/U=4$. }
\label{fc-ferri-spin}
\end{figure}

Let us introduce the vector,
\begin{equation}
 M^{\nu}_{i}= s^B \sum\limits_{\sigma\sigma'} \zeta^+_{i\sigma}
{\tau}^{\nu}_{\sigma\sigma'} \zeta_{i\sigma'}\quad {\bf M}_{i}^2=(s^B)^2 .
\label{QCB7}
\end{equation}
Then, the spin-vector of itinerant electrons Eq.(\ref{QCB4}) can be written in the
form
\be\label{QCB5a} {\bf S^B}_{i}=\frac {1}{2m}{\bf M}_{i}\left(1-h^+_i\,h_i\,-\,
d^+_i\,d_i\right),\ee
where the vector  ${\bf M}_i$ identifies the local orientation of the spin of
the sublattice B itinerant electrons.

 The Hamiltonian is quadratic with
respect to the fermions $d_i, d^+_i$ and $h_i, h^+_i$, and one can
average in the subspace of these fermions (to integrate them out in
the path integral approach). As a result, accounting for the definition of $s^B$ (see
Eq.\ref{QCB5}), one obtains ${\bf S^B}_{i}= {\bf M}_{i}$ and an effective
model for spin vectors  ${\bf S^B}_i$ with Hamiltonian
\begin{equation}\label{B1}
 H_{eff}= -J^{B2}\sum\limits_{{\ll ij \gg _B }} {{\bf S^B}_i
\cdot {\bf S^B}_j},
\end{equation}
where the effective exchange constant $J^{B2}$ is calculated in the one loop approximation. The term with $J^{B1}$ exchange constant in equation (\ref{QCB3a}) and the mixed, spin-fermion term (\ref{QCB3b}), adopt the form
\be\label{B2}
- J^{B1} \sum\limits_{{\ll ij \gg _B }} {{\bf S^B}_i \cdot {\bf S^B}_j} \ee
and
\be\label{B3}
 J\sum\limits_{  \langle  ij  \rangle } {{\bf S_i^A}}\cdot {\bf S^B_j} \ee
respectively. Collecting all terms one obtains the Hamiltonian of the FC ferrimagnetic spinel
\bea
 H & = & - J^A\sum\limits_{\ll ij \gg _A } {{\bf S}^A_{i}
\cdot {\bf S}^A_{j}}\,-\,J^B\sum\limits_{\ll ij \gg _B } {{\bf S}^B_{i}
\cdot {\bf S}^B_{j}}\nonumber \\
& & +\,J \sum\limits_{\langle ij \rangle} {{\bf S}^A_{i}
\cdot {\bf S}^B_{j}},\nonumber \eea
where $J^B=J^{B1}+J^{B2}$ and $s^B$ depending on the field applied during preparation Fig.(\ref{fc-ferri-spin}). After the process of preparation the external magnetic field is set equal to zero. For systems with $s^B=1$, discussed in the paper, one has to consider two-band model for sublattice B fermions. The result remains the same even for this more complicate system: the effective spin $s^B$ decreases when the applied, under the preparation, magnetic field increases.

\section{Hartree-Fock approximation}

Let us consider a theory with Hamiltonian (\ref{fc-ferri1}). We introduce Holstein-Primakoff representation for the spin
operators
\bea\label{rsw2} & &
S_j^{A+} = S^{A1}_j + i S^{A2}_j=\sqrt {2s^A-a^+_ja_j}\,\,\,\,a_j \nonumber \\
& & S_j^{A-} = S^{A1}_j - i S^{A2}_j=a^+_j\,\,\sqrt {2s^A-a^+_ja_j}
\\ & & S^{A3}_j = s^A - a^+_ja_j \nonumber \eea
when the sites $j$ are from sublattice $A$ and
\bea\label{rsw3} & &
S_j^{B+} = S^{B1}_j + i S^{B2}_j=-b^+_j\,\,\sqrt {2s^B-b^+_jb_j}\nonumber \\
& & S_j^{B-} = S^{B1}_j - i S^{B2}_j=-\sqrt {2s^B-b^+_jb_j}\,\,\,\,b_j
\\ & & S^{B3}_j = -s^B + b^+_jb_j \nonumber \eea
when the sites $j$ are from sublattice $B$. The operators $a^+_j,\,a_j$ and  $b^+_j,\,b_j$ satisfy the Bose commutation relations. In terms of the Bose operators and keeping only the quadratic and quartic terms, the effective Hamiltonian
Eq.(\ref{fc-ferri1}) adopts the form
\be\label{rsw4}H=H_2+H_4\ee where
\bea\label{rsw5}
 H_2 & = & s^A J^A\sum\limits_{\ll ij \gg _A }\left( a^+_i a_i\,+\,a^+_j a_j\,-\,a^+_j a_i\,-\,a^+_i a_j\right) \nonumber \\
  & + & s^B J^B\sum\limits_{\ll ij \gg _B }\left( b^+_i b_i\,+\,b^+_j b_j\,-\,b^+_j b_i\,-\,b^+_i b_j\right) \\
 & + & J \sum\limits_{\langle ij \rangle}\left[s^A b^+_j b_j + s^B a^+_i a_i -
 \sqrt{s^A s^B}\left( a^+_i b^+_j+a_i b_j \right)\right] \nonumber
 \eea
 \bea\label{rsw6}
 H_4 & = & \frac 14 J^A \sum\limits_{\ll ij \gg _A }\left[a^+_i a^+_j( a_i-a_j)^2 + (a^+_i- a^+_j)^2  a_i a_j\right] \nonumber \\
 & + & \frac 14 J^B \sum\limits_{\ll ij \gg _B }\left[b^+_i b^+_j( b_i-b_j)^2 + (b^+_i- b^+_j)^2  b_i b_j\right]\nonumber \\
 & + & \frac 14 J \sum\limits_{\langle ij \rangle}\left[
 \sqrt{\frac {s^A}{s^B}}\left( a_i b^+_j b_j b_j+a^+_i b^+_j b^+_j b_j \right)\right. \\
& + &\left. \sqrt{\frac {s^B}{s^A}}\left(a^+_i a_i a_i b_j+a^+_i a^+_i a_i b^+_j \right) - 4 a^+_i a_i b^+_j b_j \right] \nonumber
 \eea
The next step is to represent the Hamiltonian in the Hartree-Fock  approximation (\ref{fc-ferri2},\ref{fc-ferri3}) and
\bea\label{rsw9}
 H_q & = & s^A J^A u^A\sum\limits_{\ll ij \gg _A }\left( a^+_i a_i\,+\,a^+_j a_j\,-\,a^+_j a_i\,-\,a^+_i a_j\right) \nonumber \\
  & + & s^B J^B u^B\sum\limits_{\ll ij \gg _B }\left( b^+_i b_i\,+\,b^+_j b_j\,-\,b^+_j b_i\,-\,b^+_i b_j\right) \\
 & + & J u\sum\limits_{\langle ij \rangle}\left[s^A b^+_j b_j + s^B a^+_i a_i -
 \sqrt{s^A s^B}\left( a^+_i b^+_j+a_i b_j \right)\right] \nonumber
 \eea

It is convenient to rewrite the Hamiltonian in momentum space representation Eq.(\ref{fc-ferri4})
The two equivalent sublattices A and B of the bcc lattice are simple cubic lattices. The wave vector $k$ runs over the reduced  first Brillouin zone $B_r$ of a
bcc lattice which is the first Brillouin zone of a simple cubic lattice. The dispersions are given by equalities Eqs. (\ref{fc-ferri5})

To diagonalize the Hamiltonian one introduces new Bose fields
$\alpha_k,\,\alpha_k^+,\,\beta_k,\,\beta_k^+$ by means of the
transformation (\ref{fc-ferri6})
where the coefficients of the transformation $u_k$ and $v_k$ are real function of the wave vector $k$
\bea \label{ferri11} &
&u_k\,=\,\sqrt{\frac 12\,\left (\frac
{\varepsilon^a_k+\varepsilon^b_k}{\sqrt{(\varepsilon^a_k+\varepsilon^b_k)^2-4\gamma^2_k}}\,+\,1\right
)}\nonumber \\
\\
& & v_k\,=\,sign (\gamma_k)\,\sqrt{\frac 12\,\left (\frac
{\varepsilon^a_k+\varepsilon^b_k}{\sqrt{(\varepsilon^a_k+\varepsilon^b_k)^2-4\gamma^2_k}}\,-\,1\right
)}.\nonumber \eea
The transformed Hamiltonian adopts the form Eqs.(\ref{fc-ferri7})
with dispersions Eqs.(\ref{fc-ferri8})
The free energy of a system with Hamiltonian $H_{HF}$ (\ref{fc-ferri7},\ref{fc-ferri8}) is Eq.(\ref{fc-ferri10})
The system of equations for the Hartree-Fock parameters Eqs.(\ref{fc-ferri11}),
in terms of the Bose functions of the $\alpha$ and $\beta$ excitations, adopt the form
\bea\label{fc-ferri11a} u^A & = & 1-\frac {1}{3s^A} \frac 1N \sum\limits_{k\in B_r} \varepsilon_k \left[u_k^2 \,n_k^{\alpha}\, +\, v_k^2\, n_k^{\beta}\, +\, v_k^2\right]\nonumber \\
u^B & = & 1-\frac {1}{3s^B} \frac 1N \sum\limits_{k\in B_r} \varepsilon_k \left[v_k^2 \,n_k^{\alpha}\, +\, u_k^2\, n_k^{\beta}\, +\, v_k^2\right]\nonumber \\
u & = & 1-\frac 1N \sum\limits_{k\in B_r} \left[\frac {1}{2s^A}\left(u_k^2 \,n_k^{\alpha}\, +\, v_k^2\, n_k^{\beta}\, +\, v_k^2\right)\right. \\
& + & \left. \frac {1}{2s^B} \left(v_k^2 \,n_k^{\alpha}\, +\, u_k^2\, n_k^{\beta}\, +\, v_k^2\right)\right. \nonumber \\
& - & \left.8 J u\left(1+n_k^{\alpha}+n_k^{\beta}\right) \frac {\left(\cos \frac {k_x}{2} \cos \frac {k_y}{2} \cos \frac {k_z}{2} \right)^2}{\sqrt{(\varepsilon^a_k\,+\,\varepsilon^b_k)^2\,-\,4\gamma^2_k}}
\right]\nonumber
\eea
where $n_k^{\alpha}$ and $n_k^{\beta}$ are the Bose functions of $\alpha$ and $\beta$ excitations.

\section{Takahashi modified spin-wave theory \cite{Takahashi87}}

To study the paramagnetic phase of the system we make use of the Takahashi modified spin-wave theory \cite{Takahashi87} and introduce two parameters $\lambda^A$ and $\lambda^B$  to enforce the magnetization on the two sublattices to be equal to zero. It is convenient to represent the parameters
$\lambda^A$ and $\lambda^B$ in the form Eqs.(\ref{fc-ferri15}).
We find out these and Hartree-Fock parameters, as functions of temperature, solving the system of five equations, equations (\ref{fc-ferri11a}) and the equations $M^A=M^B=0$. The numerical calculations show that above N\'{e}el temperature $\mu^A\mu^B>1$ (blue line in Fig.(\ref{mumumu})). When the temperature decreases the product $\mu^A\mu^B$ decreases, remaining larger than one. The temperature at which the product becomes equal to one ($\mu^A\mu^B=1$) is the N\'{e}el temperature, marked by vertical cyan line in Fig.(\ref{mumumu}).

At low temperature $\mu^A=1$ and $\mu^B=1$ ($\lambda^A=\lambda^B=0$). The Hartree-Fock parameters are positive functions of $T/J$ , solution of the system of equations (\ref{fc-ferri11a}). Utilizing these functions, one can calculate the spontaneous magnetization on the two sublattices at low temperature. At characteristic temperature $T^*$  spontaneous magnetization on sublattice $B$ becomes equal to zero, while spontaneous magnetization on sublattice $A$ is still nonzero (see Fig.1 in the paper).

Above $T^*$ the spectrum contains long-range (magnon) excitations. Thereupon, $\mu^A\mu^B=1$ and one can use the representation
$\mu^B=1/\mu^A$. Then, $\mu^A$ and  Hartree-Fock parameters are solution of a system of four equations, equations (\ref{fc-ferri11a}) and the equation $M^B=0$.

The functions $\mu^A(T/J)$, $\mu^B(T/J)$ and $\mu^A(T/J)\mu^B(T/J)$ are depicted in figure (\ref{mumumu}) for a system with $s^A=1.5$, $s^B=1$, $J^A/J=2$ and $J^B/J=0.002$.

For the same system the Hartree-Fock parameters  $u^A(T/J)$, $u^B(T/J)$  $u(T/J)$ are depicted in figure (\ref{uuu}).

\begin{figure}[tb]
\centerline{\epsfig{width=95mm,file=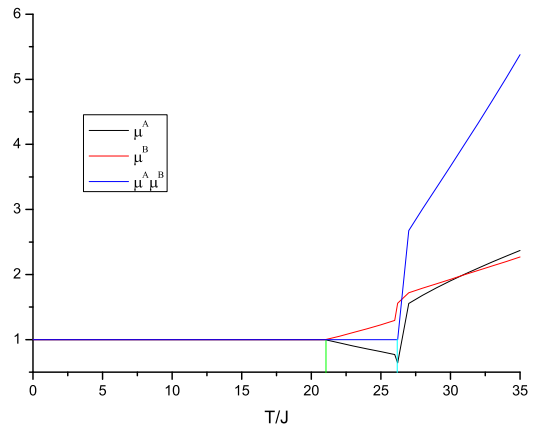}}
\caption{(Color online) Parameters $\mu^A$, $\mu^B$ and $\mu^A\mu^B$ as a function of the temperature, in units of the exchange constant $T/J$, for ferrimagnet with sublattice A spin $s^A=1.5$ and sublattice B spin $s^B=1$. The high temperature vertical (cyan) line corresponds to the N\'{e}el temperature of ferrimagnet to paramagnet transition. The low temperature vertical (green) line corresponds to partial order transition temperature.}
\label{mumumu}
\end{figure}

\begin{figure}[tb]
\centerline{\epsfig{width=95mm,file=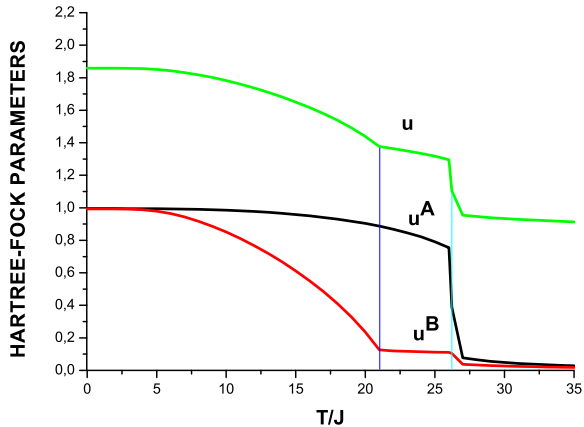}}
\caption{(Color online) Hartree-Fock parameters as a function of the temperature, in units of the exchange constant $T/J$, for ferrimagnet with sublattice A spin $s^A=1.5$ and sublattice B spin $s^B=1$. The high temperature vertical (cyan) line marks the N\'{e}el temperature of ferrimagnet to paramagnet transition. The low temperature vertical (blue) line marks the partial order transition temperature.}
\label{uuu}
\end{figure}


\begin{thebibliography}{99}

\bibitem{spinel+} K. Adachi, T. Suzuki, K. Kato, K. Osaka, M. Takata and T. Katsufuji, Phys. Rev. Lett. {\bf 95}, 197202 (2005).
\bibitem{spinelFeCr2S4} Zhaorong Yang, Shun Tan, Zhiwen Chen, and Yuheng Zhang,  Phys. Rev. {\bf B 62}, 13872 (2000).
\bibitem{spinelCv1} H. D. Zhou, J. Lu, and C. R. Wiebe, Phys. Rev. {\bf B 76}, 174403 (2007).
\bibitem{spinel08} V. O. Garlea, R. Jin, D. Mandrus, B. Roessli, Q. Huang, M. Miller, A. J. Schultz, and S. E. Nagler, Phys. Rev. Lett. {\bf 100}, 066404 (2008).
\bibitem{spinel++} S-H. Baek, K-Y. Choi, A. P. Reyes, P. L. Kuhns, N. J. Curro, V. Ramanchandran, N. S. Dalal, H. D. Zhou, and C. R. Wiebe,
J. Phys.: Condens. Matter {\bf 20}, 135218 (2008).
\bibitem{spinel11b} Kim Myung-Whun, J. S. Kim, T. Katsufuji, and R. K. Kremer, Phys. Rev. {\bf B 83}, 024403 (20011).
\bibitem{spinel11a} A. Kiswandhi, J. S. Brooks, J. Lu, J. Whalen, T. Siegrist, and H. D. Zhou, Phys. Rev. {\bf B 84}, 205138 (2011).
\bibitem{spinel11c} A. Kismarahardja, J. S. Brooks, A. Kiswandhi, K. Matsubayashi, R. Yamanaka, Y. Uwatoko, J. Whalen,
T. Siegrist, and H. D. Zhou, Phys. Rev. Lett. {\bf 106}, 056602 (2011).
\bibitem{spinel12a}Q. Zhang, K. Singh, F. Guillou, C. Simon, Y. Breard, V. Caignaert, and V. Hardy, Phys. Rev. {\bf B 85}, 054405 (2012).
\bibitem{spinel12b} Y. Nii, H. Sagayama, T. Arima, S. Aoyagi, R. Sakai, S. Maki, E. Nishibori, H. Sawa, K. Sugimoto,
H. Ohsumi, and M. Takata, Phys. Rev. {\bf B 86}, 125142 (2012).
\bibitem{spinel+1} Z. H. Huang, X. Luo, S. Lin, Y. N. Huang, L. Hu, L. Zhang, Y. P.Sun, Solid State Communications {\bf 159}, 88 (2013).
\bibitem{spinel+2} Z. H. Huang, X. Luo, L. Hu, S. G. Tan, Y. Liu, B. Yuan, J. Chen, W. H. Song, and Y. P. Sun, Journal of Applied Physics {\bf 115}, 034903 (2014).
\bibitem{spinel+3} Dina Tobia, Juli$\acute{a}$n Milano, Maria Teresa Causa and Elin L. Winkler,
J. Phys.: Condens. Matter {\bf 27}, 016003 (2015).
\bibitem{FCC-FCW} Vincent Hardy, Yohann Br$\acute{e}$ard, and Christine Martin, Phys. Rev. {\bf B 78}, 024406 (2008).
\bibitem{Diep97} R. Quartu and H. T. Diep, Phys. Rev. {\bf B 55}, 2975 (1997).
\bibitem{POexp04} J. R. Stewart, G. Ehlers, A. S. Wills, S. T. Bramwell, and J. S. Gardner, J. Phys.: Condens. Matter {\bf 16}, L321
(2004).
\bibitem{Diep87} P. Azaria, H. T. Diep, and H. Giacomini, Phys. Rev. Lett. {\bf 59}, 1629 (1987).
\bibitem{Larkin66} V. G. Vaks, A. I. Larkin, and Y. N. Ovchinnikov, JETP Letters. {\bf 22}, 820 (1966).
\bibitem{Diep04} H. T. Diep, Ed., \textit{Frustrated Spin Systems}, World Scientific (2004)
\bibitem{Takahashi87} M. Takahashi, Phys. Rev. Lett. {\bf 58}, 168 (1987).
\bibitem{Schmeltzer} D. Schmeltzer, Phys. Rev. B {\bf 43}, 8650 (1991).

\bibitem[*]{byline} Electronic address: naoum@phys.uni-sofia.bg
\end{thebibliography}
\end{document}